\documentclass[12pt]{article}
\usepackage[margin=2 cm]{geometry}
\usepackage{comment}
\usepackage{url}
\usepackage{color}
\usepackage{graphicx}
\usepackage[font={small,it}]{caption}
\usepackage{mwe}
\usepackage[nottoc]{tocbibind}
\usepackage{amsmath,amssymb,extarrows,mathtools,graphicx,subfigure,setspace}
\usepackage{cite}
\usepackage{tikz}\usetikzlibrary{calc}
\usepackage{braket}
\usepackage{epsfig}
\usepackage[section]{placeins}
\usepackage{slashed}
\usepackage{color}
\usepackage{caption}
\usepackage{amsmath}
\usepackage{hyperref}
\makeatother

\newcommand{\be}{\begin{equation}}
\newcommand{\bea}{\begin{eqnarray}}
\newcommand{\eea}{\end{eqnarray}}
\newcommand{\ba}{\begin{array}}
\newcommand{\ea}{\end{array}}
\newcommand{\ee}{\end{equation}}
\newcommand{\bes}{\begin{equation*}}
\newcommand{\beas}{\begin{eqnarray*}}
\newcommand{\eeas}{\end{eqnarray*}}
\newcommand{\bas}{\begin{array*}}
\newcommand{\eas}{\end{array*}}
\newcommand{\ees}{\end{equation*}}

\numberwithin{equation}{section}

\begin{document}

\onehalfspacing
\vfill
\begin{titlepage}
\vspace{10mm}

\begin{center}

\vspace*{10mm}
\vspace*{1mm}
{\Large  \textbf{Critical distance and Crofton form in confining geometries}} 
 \vspace*{1cm}
 
{$\text{Mahdis Ghodrati}^{a}$},

\vspace*{8mm}
{ \textsl{
$^a $Asia Pacific Center for Theoretical Physics, Pohang 37673, Republic of Korea}} 
 \vspace*{0.4cm}

\textsl{e-mails: {\href{mahdis.ghodrati@apctp.org}{mahdis.ghodrati@apctp.org}}}
 \vspace*{2mm}

\vspace*{1.7cm}

\end{center}

\begin{abstract}
For two symmetric strips with equal and finite size and in the background of several confining geometries, we numerically calculate the critical distance between these two mixed systems where the mutual information between them drops to zero and show that this quantity could be a useful correlation measure in probing the phase structures of holographic QCD models. The models that we consider here are Sakai-Sugimoto and deformed Sakai-Sugimoto, Klebanov-Tseytlin and Maldacena-Nunez. For evaluating the structures of these holographic supergravity geometries from the perspective of the bulk reconstruction, we also calculate their Crofton forms and show that there is a universal behavior in the confining backgrounds where a \textit{``well functionality"} is present around the IR cutoff point, and far from the IR wall the scalar part of the Crofton form would become constant, demonstrating the effects of the wall of the confining models on the phase structures. This work is the shorter version of our previous work arXiv:2110.12970 with few more results about the connections between phases.
 \end{abstract}

\end{titlepage}

\tableofcontents


\section{Introduction}

The Ryu-Takayanagi (RT) prescription \cite{Ryu:2006bv, Ryu:2006ef} in holography \cite{Witten:1998qj} gives a simple method for calculating the entanglement entropy of field theories \cite{Casini:2009sr} by using the calculations of the area of the extremal surface which is homologous to the boundary. This idea showed the deep connections between quantum information and geometry. Other new holographic dualities in this context such as computational complexity/volume of the subregions \cite{Alishahiha:2015rta,Brown:2015bva,Ghodrati:2017roz,Ghodrati:2018hss,Ghodrati:2019bzz,Zhou:2019jlh}, and entanglement of purification(EoP)/minimal wedge cross section \cite{Takayanagi:2017knl,Ghodrati:2019hnn,Ghodrati:2020vzm} further strengthened this interrelation.

So as the calculations of the properties of geometrical objects in the bulk are much easier than the calculations of the quantum measures in the boundary field theories and some of these geometrical objects can probe the deep bulk structures, one would expect that calculating these quantities in various models can also detect the heterogeneous phase structures of the exotic forms of boundary systems, for instance in models such as holographic QCD or AdS/CMT.

The motivation of this work is, in the background of confining geometries, using the critical distance between two mixed systems coming from the change of the orders in their mutual information, we depict the phase structures of several holographic QCD models. We show that, this property can also disclose the scale of chiral symmetry breaking versus the scale of confining-deconfining phase transition. Additionally, in any holographic QCD model, one would have an IR wall where the geometry would be terminated in the holographic direction. In this work we also check the effects of these walls or the terminated radius $u_{KK}$ (or $r_s$) on the quantum information measures. So we show that the relative scales in the system, i.e, the distance between the two strips, the width of the strips and the $u_{KK}$ would lead to different phases, and then we further examine the effects of $u_{KK}$ by studying the Crofton form.

So, first in section \ref{sec:generalize}, we present the extended relations for correlation measures in the confining backgrounds, and the results for each of our confining models.
Then, in section \ref{sec:Crofton}, for understanding better the structures of these top-down models, we present the behavior of Crofton-form \cite{Czech:2015qta} and show that in all of these confining cases there is a universal behavior where a \textit{``well shape"} can be detected around the IR wall, while its absolute value becomes constant at larger holographic radius, which again demonstrates the effects of $u_{KK}$ point on the correlation measures and the phase diagrams in these models. We then conclude in section \ref{sec:conclude}.

\section{Critical distance in confining backgrounds}\label{sec:generalize}
The generalization of entanglement entropy (EE) would be from $S_A= \frac{1}{4 G_N^{(d+2)} } \int_\gamma d^d \sigma \sqrt{G^{(d)}_{\text{ind}} }$ for the conformal case, to the relation $S_A=\frac{1}{4G_N ^{(10)}} \int d^8 \sigma e^{-2 \phi} \sqrt{G^{(8)} _{\text{ind}}}$, for the confining models, where $G_N^{(d+2)}$ is the $d+2$ dimensional Newton constant and $G_{\text{ind} }^{(d)}$ is the induced string frame metric on $\gamma$. This generalization would be due to the fact that, in the non-conformal theories, the volume of the $8-d$ compact dimensions and also the dilaton are not constant.

In the background of confining models we consider two strips with equal width, and by changing the three main scales in the system relative to each other, where they are $D$, the distance between the two strips, $L$, the width of the strips and $r_{s}$ (or $u_{KK}$), the position of the wall in the confining geometry, we show one can probe various phases of these confining backgrounds.

Any confining gravitational backgrounds in the string frame could be written in the form of \cite{Kol:2014nqa}
\begin{gather}
ds^2= \alpha( \rho) \lbrack \beta(\rho) d\rho^2 + dx^\mu dx_\mu \rbrack + g_{ij} d \theta^i d \theta^j,
\end{gather}
where $\rho$ is the holographic radial coordinate in the interval of $\rho_\Lambda < \rho < \infty $,  and $x^\mu, \ \ (\mu=0,1,...,d)$ parameterizes  the $\mathbb{R}^{d+1}$, and $\theta^i, \ \ (i= d+2, . . . ,9)$ are the $8-d$ internal directions, where our calculations would be done in this frame.

On the other hand, the bulk minimal wedge cross section between the two mixed boundary regions, as a measure of correlation, has been proposed in \cite{Takayanagi:2017knl} to be dual to the entanglement of purification between the systems of $A$ and $B$. For the simple Schwarzchild AdS black brane geometry and for two regions with width of $l$ and distance $D$, this quantity has been found as $\Gamma=\int_{z_D}^{z_{2l+D}} \frac{dz}{ z^{d-1} \sqrt{1-\frac{z^d}{z_h^d} }}$, where $z_D$ and $z_{2l+D}$ are the two turning points. This then can be generalized to the case of confining geometries. For Dp-brane geometries, for instance, in \cite{Lala:2020lcp}, several similar quantum information measures have been calculated.

Here we could generalize it as 
\begin{gather}\label{eq:generalizedeEW}
\Gamma=\int_{u_D}^{u_{2l+D} } du e^{-2 \Phi} \sqrt{-\gamma_{ab} },
\end{gather}
where $\gamma_{ab}$ is the determinant of the eight dimensional induced metric on $\Gamma$ that comes from the ten dimensional metric.

Also, $e^{-2\Phi}$ should be considered, since in the non-conformal geometries, the dilaton field $\Phi$ and the volume of the compact dimensions are not constant and so similar to the entanglement entropy case, are needed to be inserted in the relation for the measure. 

Defining the quantity $H(\rho) = e^{-4 \phi } V^2_{\text{int}} \alpha^2$, where $V_{\text{int}} = \int d \vec{\theta} \sqrt{\text{det} \lbrack g_{ij} \rbrack }$ is the volume of the internal manifold described by the coordinates $\vec{\theta}$, the entanglement entropy (EE) for the connected solution $S_C$ and for the disconnected solution $S_D$ could be written as \cite{Kol:2014nqa}
\begin{equation} \label{eq:condisL}
\begin{split}
S_C(\rho_0) &= \frac{V_{d-1} }{2 G_N^{(10)} } \int_{\rho_0}^\infty d\rho \sqrt{ \frac{\beta(\rho) H(\rho) }{1- \frac{H(\rho_0) }{H(\rho) } }}, \ \ \ \  S_D (\rho_0) = \frac{V_{d-1} }{2 G_N^{(10)} } \int_{\rho_\Lambda}^\infty d\rho \sqrt{\beta(\rho) H(\rho)},
\end{split}
\end{equation}
and the width of one interval as a function of $\rho_0$, which is the minimum value of the radial coordinate, could be written as $L(\rho_0) =2 \int_{\rho_0}^ \infty d\rho \sqrt{\frac{\beta(\rho) }{ \frac{H(\rho)}{H(\rho_0)}-1} }$.

Note that these quantities are defined only for the pure states. When we have mixed system, the main correlation measure is the mutual information, $I(A,B)= S_A + S_B - S_{AB}$, which by inserting the relation for the connected solution of \ref{eq:condisL} in it could be extended to the confining models. From the width of one interval $L$ and the distance between intervals $D$, the turning point for each region and then their entanglement entropies can be calculated. In addition, from the distance between the turning points, the minimal wedge cross section $\Gamma$ can be found.

The interesting point is that when the distance between two mixed systems becomes larger, at a specific critical distance $D_c$, the correlations, mutual information and entanglement of purification between them vanish or rather drop to one order of magnitude lower. For two equal strips with the width $l$ and distance $D$, the mutual information is
\begin{gather}\label{eq:mutualrel}
I(D,l)= S_A+S_B - S_{AB}= 2S(l)-S(D)-S(2l+D),
\end{gather}
and the critical distance is found by where it drops to zero, i.e, $I(D_c,l)=0$. We then show that this critical distance could give interesting information about the QCD phase structure.

In fact, even the critical width for ``one strip'' could distinguish the confinement/deconfinement phase transition and it has been proposed that this critical width $L_{\text{crit}}$ is proportional to the inverse of confinement temperature, i.e, $L_{\text{crit} }= \mathcal{O} (\Lambda_{IR}^{-1})$, where $\Lambda_{IR}$ correspond to the IR cut-off or IR wall. Here we can show that the critical distance between \textit{``two strips''} is rather a stronger quantity which can even distinguish the scale of chiral symmetry breaking in addition to  the confinement/deconfinement scale.

\begin{figure}[ht!]   
\begin{center}
\includegraphics[width=0.99\textwidth]{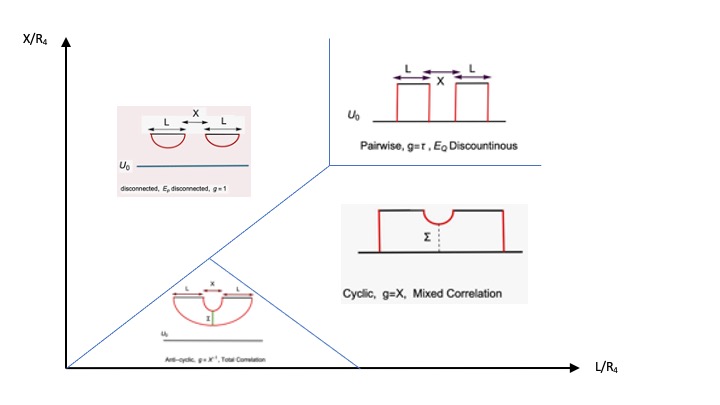}  
\caption{Phase diagram for confining geometries using mutual information and $D_c$}
\label{fig:parulphases}
\end{center}
\end{figure}

Similar to the work of \cite{Jain:2020rbb}, coming from negativity, we also find three jumps in the behavior of $D_c$, and so the phase diagram again has four different phases in these confining models as shown in figure \ref{fig:parulphases}.

\begin{figure}[ht!]   
\begin{center}
\includegraphics[width=0.65\textwidth]{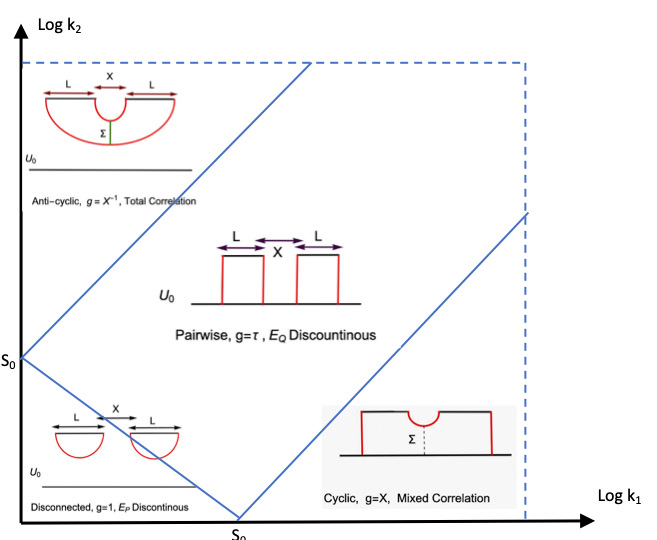}  
\caption{Phase diagram of mixed system using negativity from \cite{Dong:2021oad} and comparing with plots of $D_c$ in confining geometries.}
\label{fig:fphasesNeg}
\end{center}
\end{figure}

\begin{figure}[ht!]   
\begin{center}
\includegraphics[width=0.8\textwidth]{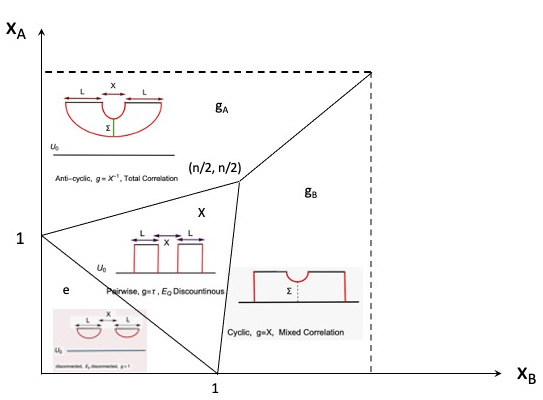}  
\caption{Phase diagram of mixed system using wormholes \cite{Akers:2021pvd} and comparing with plots of $D_c$ in confining geometries. }
\label{fig:Ackersdiagram}
\end{center}
\end{figure}

\begin{figure}[ht!]   
\begin{center}
\includegraphics[width=0.8\textwidth]{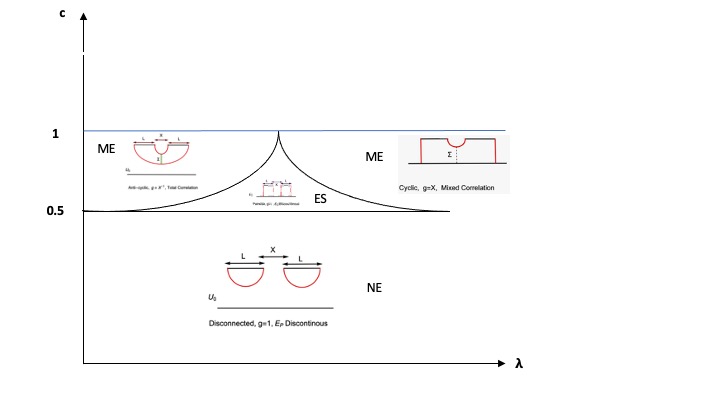}  
\caption{Phase diagram using negativity derived in \cite{Vardhan:2021npf,Shapourian:2020mkc}. }
\label{fig:shapur}
\end{center}
\end{figure}

Note that in addition to mutual information and critical distance which we have investigated here to derive the phase structures of confining geometries with a wall, one could use other mixed correlation measures such as logarithmic negativity, reflected entropy, entanglement of formation, distillable entanglement, squashed entanglement, relative entropy of entanglement, etc, to find the phase structures of mixed systems in various setups, and then compare them to find the universal properties of quantum and classical correlations. Several phase diagrams which have been recently found using mixed correlation measures have been shown in figures \ref{fig:fphasesNeg}, \ref{fig:Ackersdiagram}, and \ref{fig:shapur} here and they have been compared with the phases derived for two symmetric mixed strips. In future works, we report further on the connections between these phase diagrams and universal intrinsic properties within the correlation of Hawking radiation of black holes, and information paradox.

\subsection{Sakai-Sugimoto and deformed Sakai-Sugimoto}
\begin{figure}[ht!]
 \centering
    \includegraphics[width=7 cm] {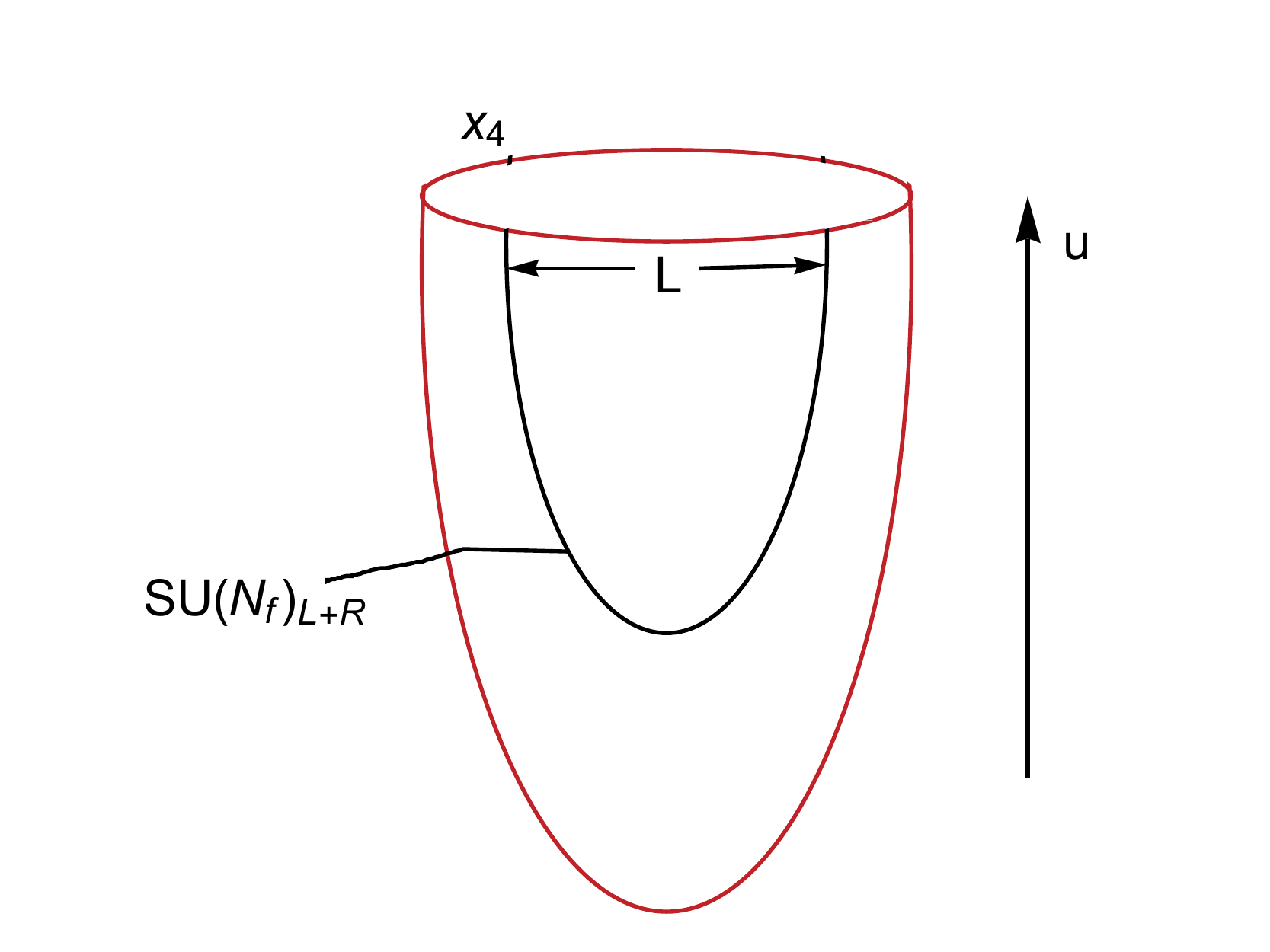}
  \includegraphics[width=7 cm] {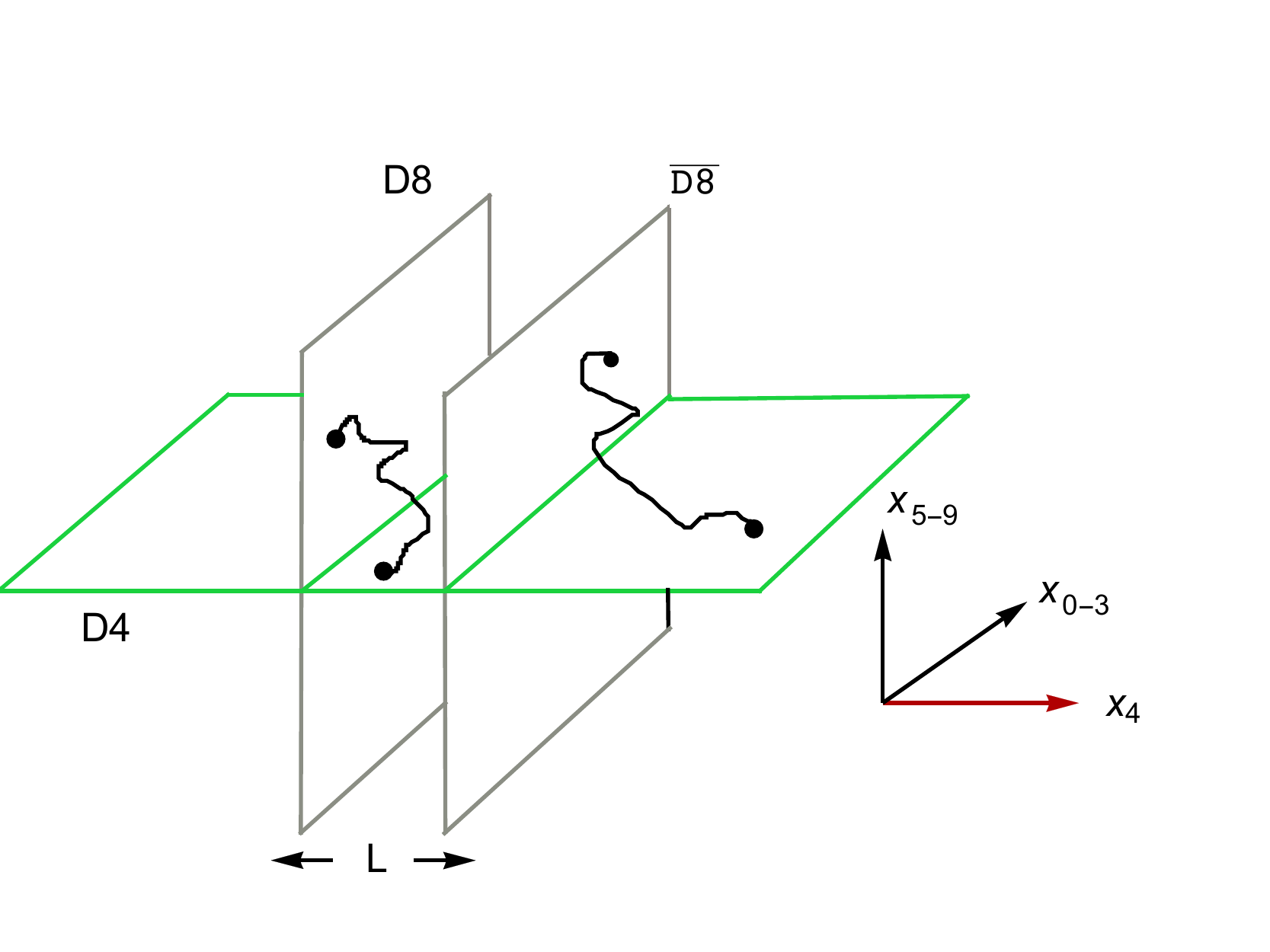}
  \caption{Intersecting D4-D8 brane model of Sakai-Sugimoto. Usually the D8-branes are treated as probe branes.}
 \label{fig:sakai}
\end{figure}

In this work we are interested in the top-down holographic QCD models which are engineered by intersecting D-branes, like D4-D8 brane intersections known as Sakai-Sugimoto \cite{Sakai:2004cn,Sakai:2005yt}  model shown in figure \ref{fig:sakai}. This theory can nicely model the $SU(N_f)_L \times SU(N_f)_R$ chiral flavor symmetry breaking.

In this setup, $x^4$ is the spatial coordinate which is being compactified on $S^1$, which for fermions would have anti-periodic boundary conditions. The D8-branes are at $x^4=0$ and the anti-D8 branes are located in parallel and at $x^4=\pi R$, and $R$ is the radius of $S^1$.

The entanglement and correlations structures among D8-D8, D4-D4 and D8-D4 branes would be different, where it can show their effect on the critical distance $D_c$.

The metric of D4-branes is \cite{Brandhuber:1999np}
\begin{gather}
ds^2_{D4}= \left(\frac{u}{R_{D4}} \right)^{3/2} (-dt^2 +\delta_{ij} dx^i dx^j +f(u) (dx^4)^2 ) + \left(\frac{R_{D4} }{u}\right)^{3/2} \left( \frac{du^2}{f(u)} + u^2 d \Omega_4^2 \right ),
\end{gather}
where the dilaton, the Ramond-Ramond field, the function $f(u)$ and the AdS radius are defined as
\begin{gather}
e^\phi= g_s  \left(\frac{u}{R_{D4}} \right)^{3/4}, \ \ F_4\equiv dC_3= \frac{2\pi N_c}{V_4} \epsilon_4, \ \ f(u) \equiv 1-\frac{u_{KK}^3 }{u^3}, \ \ R^3_{D4} \equiv \pi g_s N_c l_s^3.
\end{gather}
Here, $N_c$ is the number of colors in the gauge group, $g_s$ is the string coupling, $l_s$ is the string length which has the relation $l_s^2= \alpha^\prime$, $V_4$ is the volume of the unit four sphere $S^4$, and $s_4$ is the volume form of $S^4$. In addition, $u$ is the holographic radial direction, which is in the region of $u_{KK} \le u \le \infty$.

Then, assuming the position of D8 brane  \cite{Brandhuber:1999np, Brodsky:2014yha} at $x^4=0$ and anti-brane at $x^4=\pi R$, leading to relation $dx^4/du=0$, the induced metric of the D8-brane becomes
\begin{gather}\label{eq:metricd8}
ds_{D8}^2 = \left (\frac{u}{R_{D4} }  \right)^{3/2} (-dt^2+\delta_{ij} dx^i dx^j )+ \left (\frac{R_{D4} }{u}  \right)^{3/2} \left ( \frac{du^2}{f(u) } +u^2 d\Omega_4^2 \right). 
\end{gather}

The calculations here should be done for the flavor D-branes metric which represent the quark sector, which here for the case of Sakai-Sugimoto, would correspond to the \textit{D8-branes}. In works such as \cite{Hashimoto:2014yya}, also for calculating the pair production rates, the imaginary part of D8-branes have been evaluated. Therefore, for the calculation of Sakai-Sugimoto case, we should consider just the above metric \ref{eq:metricd8}.

For this metric, we find $S_C(u_t)$, the entanglement entropy at turning point $u_t$ of the strip with width $L(u_t)$ and the width of the strip in terms of turning point as
\begin{gather}\label{eq:SSakai}
S_C(u_t)=\frac{V_3 V_4 R_{D_4}^3 }{2 g_s^2 G_N^{(10)}}  \int_{u_t}^\infty du  \frac{u }{\sqrt{ \left(1-\frac{u_{KK}^3 }{u^3} \right)  \left(   1-  \frac{u_t^5}{u^5} \right )  } }, \nonumber\\
L(u_t)= 2 R_{D_4}^{\frac{3}{2}} \int_{u_t}^ \infty du \frac{1}{\sqrt{u^3 \left(1- \frac{u_{KK}^3 }{u^3} \right)  \left( \frac{u^5}{u_t^5} -1 \right)   } }. 
\end{gather}

\begin{figure}[ht!]
 \centering
  \includegraphics[width=8.5 cm] {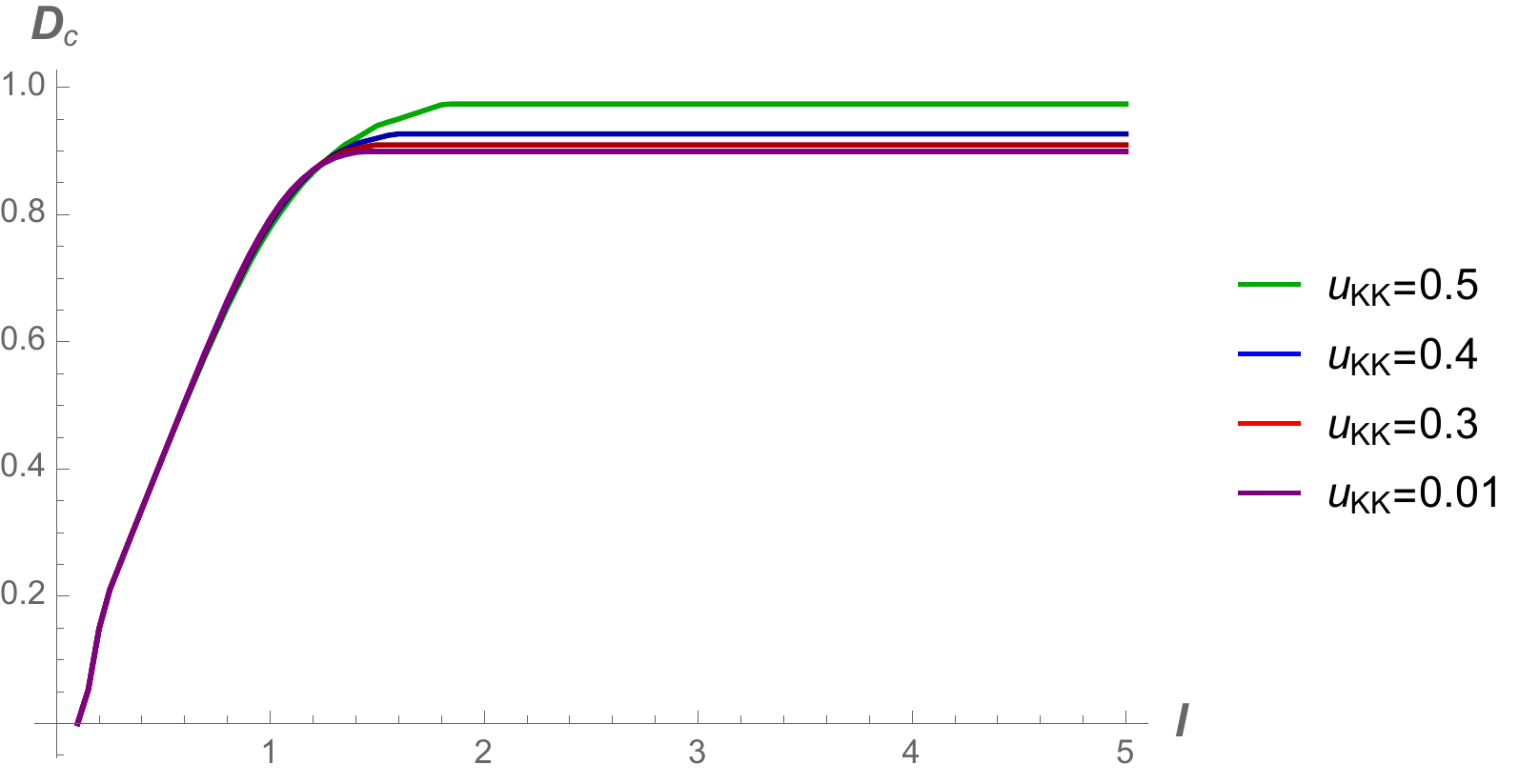} 
  \caption{The plot of $D_c$ vs. $l$ for different $u_{KK}$ in phase 1.}
 \label{fig:Dc1sakaisugimoto}
\end{figure}

\begin{figure}[ht!]
 \centering
  \includegraphics[width=8.5 cm] {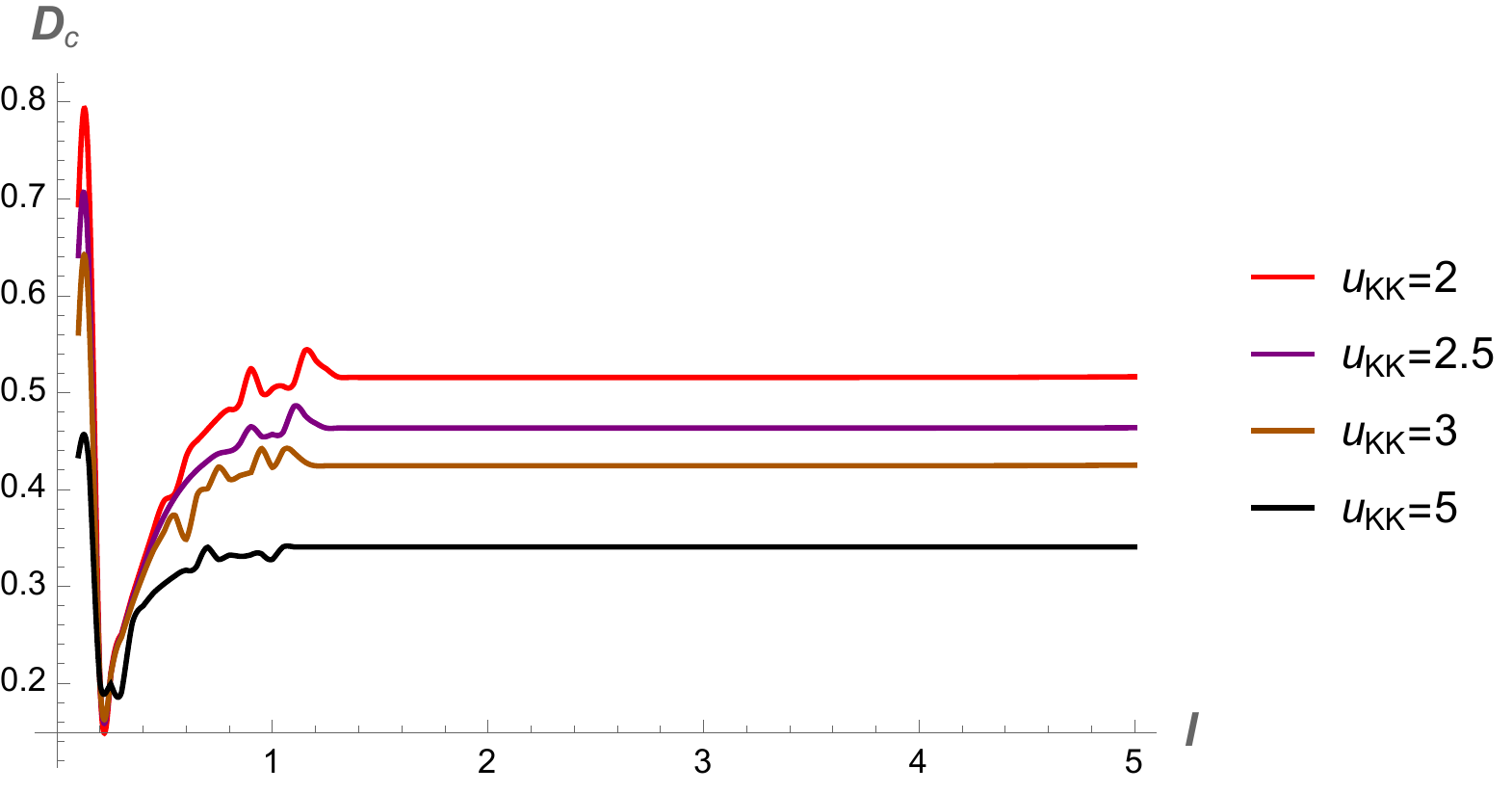} 
  \caption{The plot of $D_c$ vs. $l$ for different $u_{KK}$ in phase 2.}
 \label{fig:DPhase2Sakai}
\end{figure}

\begin{figure}[ht!]
 \centering
\includegraphics[width=8.5 cm] {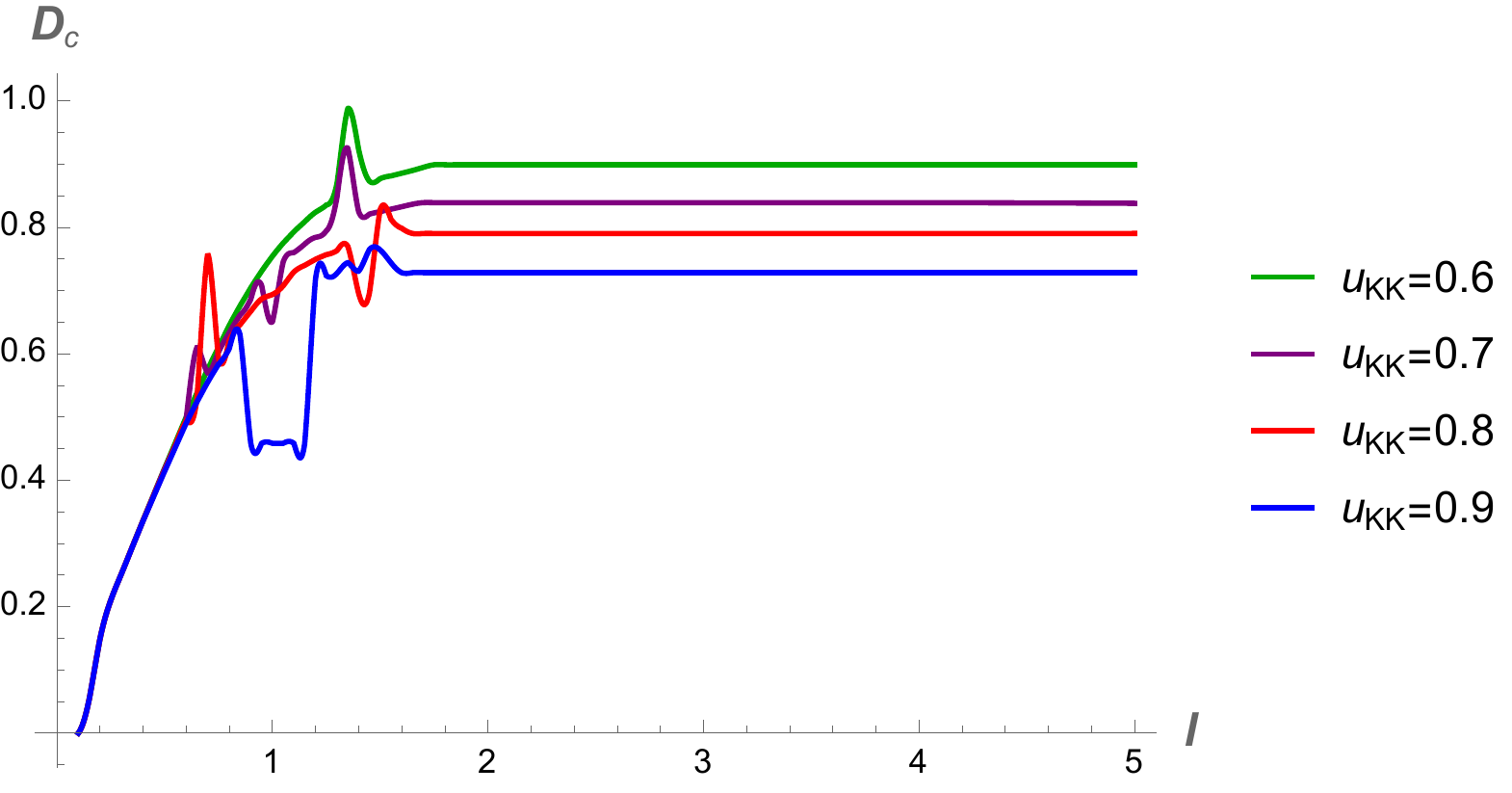} 
  \caption{The plot of $D_c$ vs. $l$ for different $u_{KK}$ in phase 3.}
 \label{fig:DPhase3Sakai}
\end{figure}

Next, in this confining model, we examine the behavior of critical distance between the two strips $D_c$ as a function of $u_{KK}$ and detect various phases in this specific background as shown in figures \ref{fig:Dc1sakaisugimoto}, \ref{fig:DPhase2Sakai}, \ref{fig:DPhase3Sakai} .

As we expected, four distinct phases could be detected. In the first phase shown in figure \ref{fig:Dc1sakaisugimoto}, increasing $u_{KK}$ would increase $D_c$. When $u_{KK}$ becomes zero the critical distance would be around $D_c=0.899$. The upper bound for $u_{KK}$ in this phase is $u_{KK}=0.5$ corresponding to $D_c= 0.97$. After that, increasing $u_{KK}$ would decrease $D_c$ and we enter the next phase shown in figure \ref{fig:DPhase2Sakai}.

In the third phase shown in figure \ref{fig:DPhase3Sakai}, again one can see that after reaching its maximum at the previous phase, increasing $u_{KK}$ decreases $D_c$.  In the fourth phase however, the behavior of $D_c$ in this model would be constant and zero which we did not show here.

\begin{figure}[ht!]
 \centering
  \includegraphics[width=10 cm] {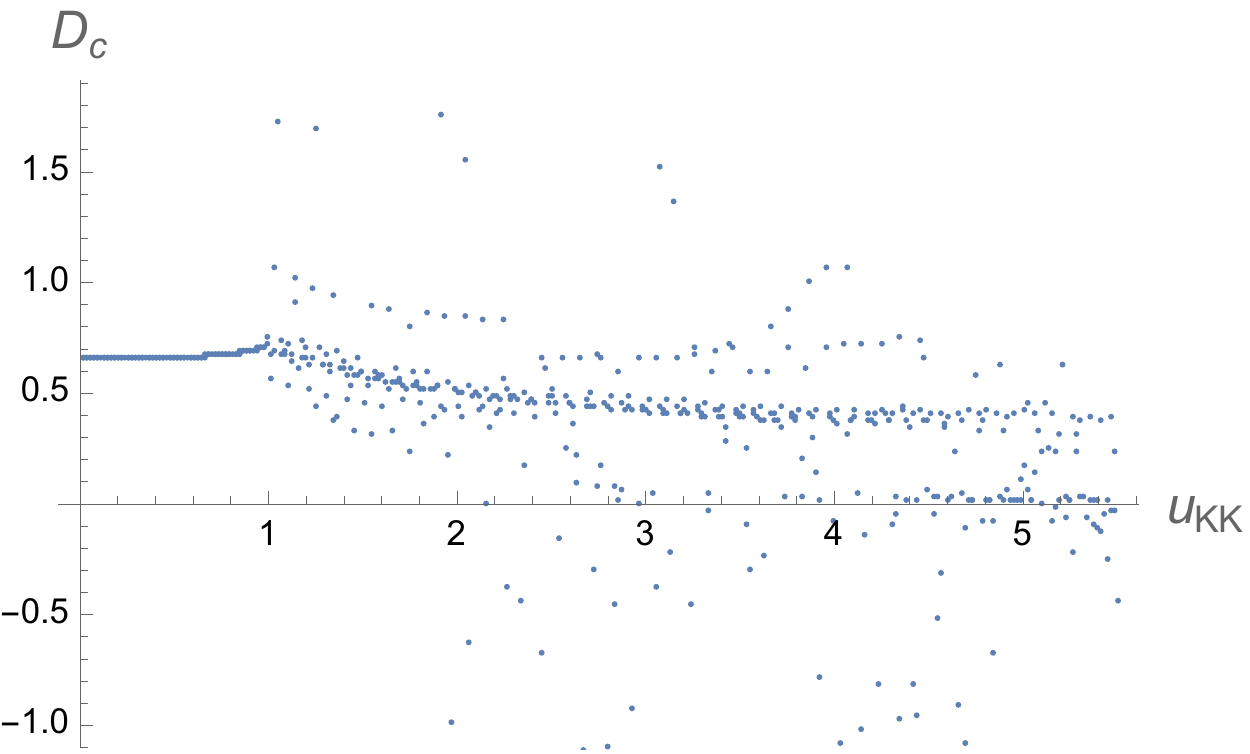} 
  \caption{The plot of $D_c$ vs. $u_{KK}$ in Witten-Sakai-Sugimoto background.}
  \label{fig:dotplot2}
\end{figure}

 One of these phase transitions is related to chiral symmetry breaking/restoration and the other sharp phase transition is due to the confinement/deconfinement transition related to the breaking/restoring of $Z_{N_c}$ global symmetry and the last one is due to the change of the order in mutual information or when the entanglement wedge becomes disconnected. Note that as found in \cite{Aharony:2006da}, the chiral symmetry would be broken in the confined phase, therefore, the first phase transition seen in figure \ref{fig:dotplot2} is due to the chiral symmetry breaking and the second one should be due to the deconfinement.

These phase transitions could also be examined from the point of view of traversable wormholes \cite{Maldacena:2018lmt, Numasawa:2020sty}. During these first order phase transitions,  the traversable wormholes would exchange their configurations and saddles. During the phase transitions between these saddles generally the resolvent function for calculating the mixed measures such as negativity would also diverge, check figure 5 of \cite{Dong:2021oad} for instance, which is similar to the behavior of $D_C$ here. This is because at the phase transition all the saddles contribute and therefore $D_C$ could increase while keeping the mutual information non-zero.

So the competitions between chiral symmetry breaking/restoration and the vanishing of the mutual information due to increasing the distance between the two strips would create these four phases. This is because the scale of chiral symmetry breaking and confinement are independent, as the scale of chiral symmetry breaking mostly depends on the distance between D8-branes or the fermions on the compactified circle, while the scale of confinement would depend on the radius where D4-brane is wrapped around, or the radius of the circle which the $SU(N_c)$ gauge theory is compactified on it.

Then,  the relation for the minimal wedge cross section in this model would be
\begin{gather}
\Gamma_{WSS}= R_{D4}^3 \int_{u_D}^{u_{2l+D}} du \frac{  u ^5 }{1-\frac{u^3_{KK} }{u^3} }.
\end{gather}

Based on the AdS/CFT dictionary there are connections between the parameters $M_{KK}, g_{YM}, N_c$ in the boundary gauge side and $R_{D_4}, u_{KK}, g_s$ in the gravity side as
\begin{gather}
u_{KK}=\frac{2}{9} \lambda M_{KK} l_s^2, \ \ \ g_s=\frac{1}{2\pi} \frac{\lambda}{M_{KK} N_c l_s  }, \ \ \ \ R_{D_4}^3=\frac{1}{2} \frac{\lambda l_s^2}{M_{KK}}.
\end{gather}

The 't Hooft coupling $\lambda$ has the relation $\lambda= g_{YM}^2 N_c $ and the gauge coupling $g_{YM}$ at the cutoff scale $M_{KK}$ can be written as $g_{YM}^2=(2\pi)^2 g_s l_s/ \delta x^4 $ and we also have the relation $M_{KK}\equiv \frac{2\pi}{\delta x^4}  $\cite{Hashimoto:2014yya}.

\begin{figure}[ht!]
 \centering
  \includegraphics[width=8 cm] {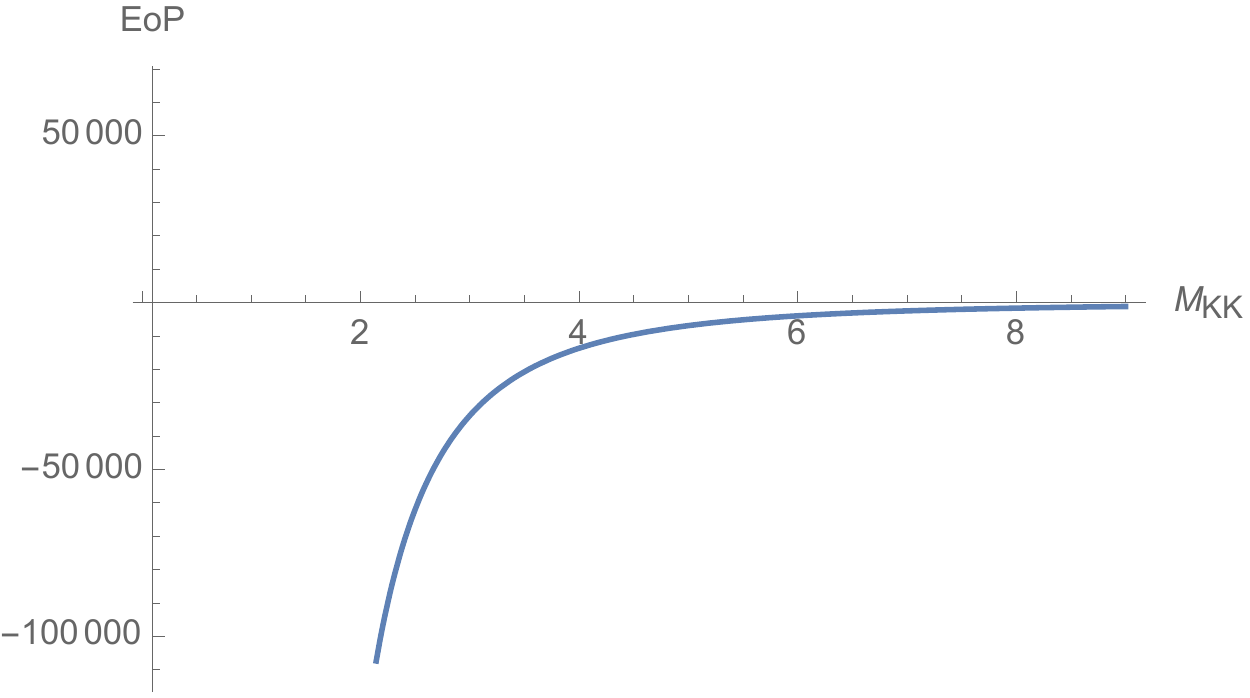} 
    \includegraphics[width=8 cm] {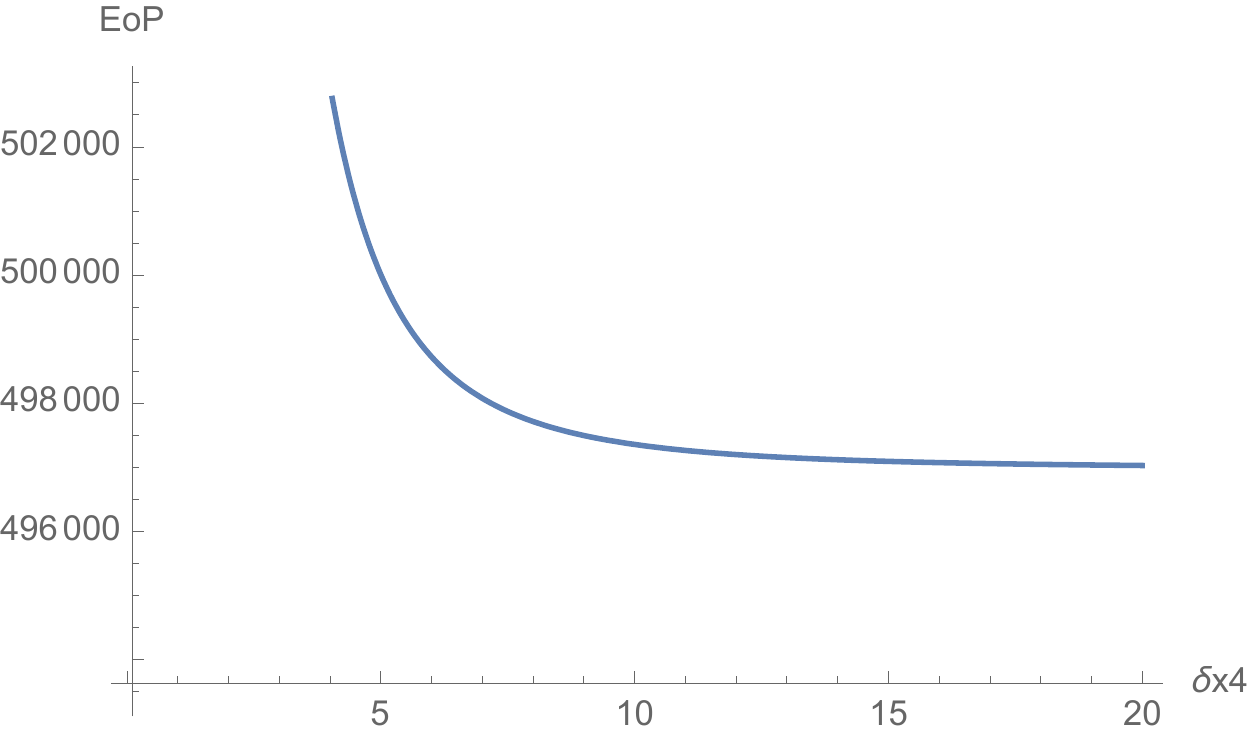} 
  \caption{The behavior of entanglement of purification versus $M_{KK}$ and $\delta x_4$ }
  \label{fig:dEoPMKK}
  \end{figure}

So, it could be seen that the string coupling $g_s$, the number of colors of gauge group $N_c$, the string length $l_s$ and the cutoff energy $M_{KK}$ increase the entanglement of purification while the periodicity of the boundary condition $\delta x^4$ would have an inverse effect and decreases the EoP. The behavior of EoP versus $M_{KK}$ and its behavior versus $\delta x^4$ are shown in figure \ref{fig:dEoPMKK}. 

One can then consider the deformed Sakai-Sugimoto geometry \cite{Aharony:2006da,Hashimoto:2014yya}, where $x^4$ instead of being constant, depends on the coordinate $u$. 

\begin{figure}[ht!]
 \centering
  \includegraphics[width=7 cm] {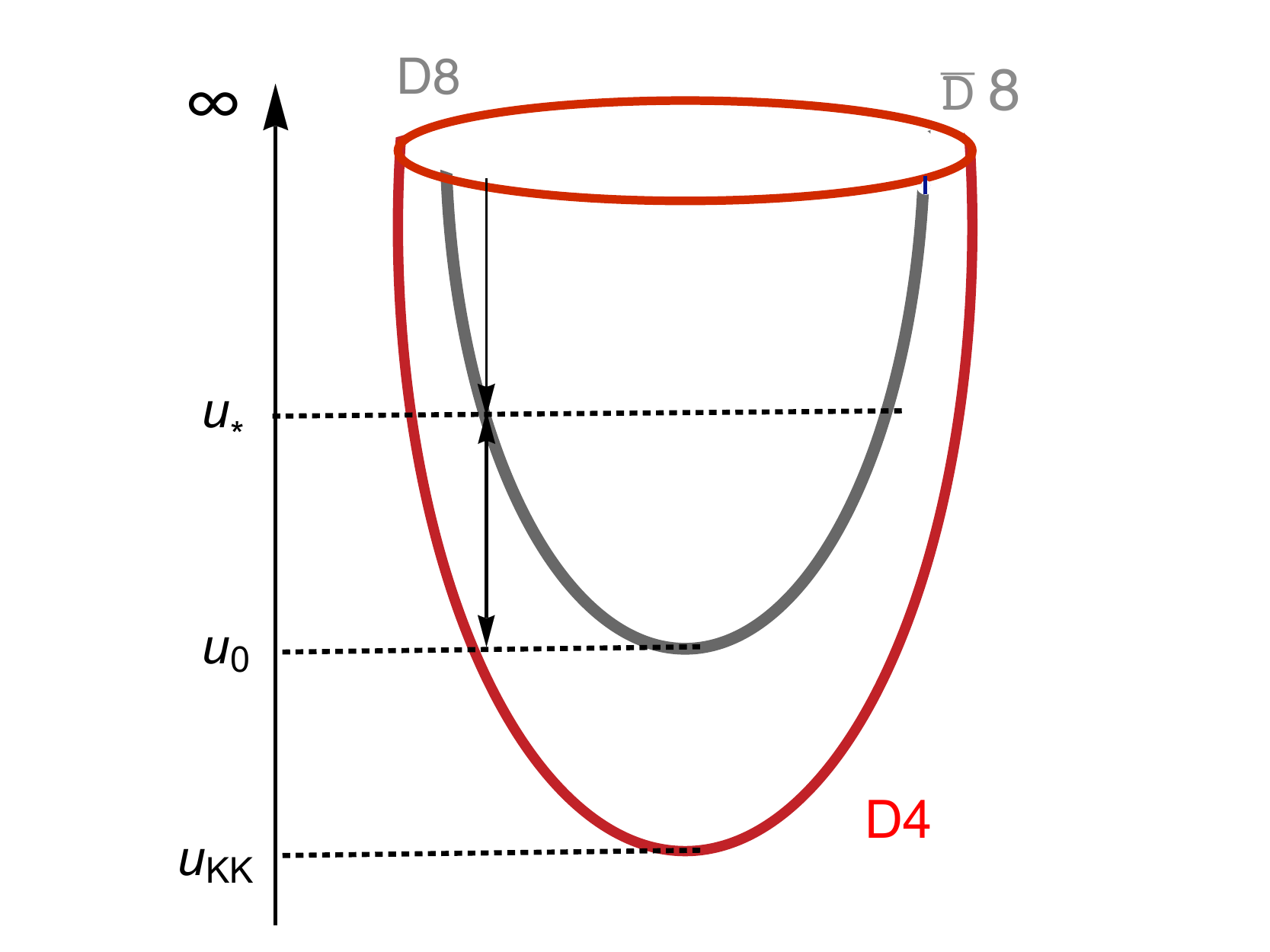} 
  \caption{The geometry of deformed Sakai-Sugimoto.}
  \label{fig:deformedSakai}
  \end{figure}

The deformed Sakai-Sugimoto model \cite{Hashimoto:2014yya}. has the configuration shown in figure \ref{fig:deformedSakai} and its metric is 
\begin{gather}
ds^2_{D8}= \left ( \frac{u}{R_{D4}} \right)^{3/2} (-dt^2+\delta_{ij} dx^i dx^j ) +\left ( \frac{u}{R_{D4}} \right)^{3/2} \frac{du^2}{h(u)} +\left (\frac{R_{D4} }{u} \right)^{3/2} u^2 d \Omega_4^2,
\end{gather}
where 
\begin{gather}
h(u) \equiv \left \lbrack f (u) \left ( \frac{dx^4 (u) }{du} \right )^2 + \left ( \frac{R_{D4} }{u} \right)^3 \frac{1}{f(u)} \right \rbrack ^{-1}.
\end{gather}

The entanglement entropy of a strip in terms of the turning point could be written as
\begin{gather}
S(u_t)=\frac{V_3 V_4 R_{D_4}^3 }{2 g_s^2 G_{N}^{(10) }  } \int_{u_t} ^\infty du \frac{u}{\sqrt { (1-\frac{u_{KK}^3 }{u^3} ) (1-\frac{u_t^5}{u^5} ) } } + \frac{V_3 V_4 R^{\frac{3}{2} }_{D_4} }{2 g_s^2 G_N^{(10) } } \int_{u_t}^ \infty du \left( \frac{dx^4 (u)}{du}  \right) \sqrt{ \frac{u^5 (1-\frac{u_{KK}^3 }{u^3} ) }{1-\frac{u_t^5}{u^5} }  }.  
\end{gather}

Similarly, the function of width of the strip in terms of the turning point $u_t$ is
\begin{gather}
L(u_t)= 2 R_{D_4}^{\frac{3}{2}} \int_{u_t}^ \infty du \frac{1}{\sqrt{u^3 \left(1- \frac{u_{KK}^3 }{u^3} \right)  \left( \frac{u^5}{u_t^5} -1 \right)   } }+2\int_{u_t}^\infty du \left( \frac{dx^4 (u) }{du} \right)  \sqrt{\frac{1-\frac{u_{KK}^3 }{u^3} }{\frac{u^5}{u_t^5} -1 } },
\end{gather}
which just the second term has been added to the previous relation.

For this case then, the relation for the minimal wedge cross section would be
\begin{gather}
\Gamma_{\text{Deformed SS}}=  R_{D4}^3 \int_{u_D}^{u_{2l+D}} du \frac{  u ^5 }{1-\frac{u^3_{KK} }{u^3} } +\int_{u_D}^{u_{2l+D}} du u^8 \left(1- \frac{u^3_{KK} }{u^3} \right)  \left(\frac{dx^4(u)}{du} \right)^2.
\end{gather}

The difference between the minimal wedge cross section of deformed Sakai-Sugimoto and un-deformed case is being controlled by the derivative of $x^4(u)$ relative to $u$. 

By calculating the DBI action and then calculating its Hamiltonian which is conserved one would specifically get the relation for $u' =du/dx^4$ \cite{Aharony:2006da} as $u'^2= f^2 (u) \left ( \frac{u}{R_{D_4} } \right)^3 \left ( \frac{f(u)}{f(u_0)} \frac{u^8}{u_0^8}-1 \right )$. Then, we can insert the inverse of this relation into all the previous results for the deformed Sakai-Sugimo and find the numerical solutions, which again gives the same three phase transitions corresponding to the four phases.

\subsection{Klebanov-Tseytlin}
The Klebanov-Tseytin (KT) metric \cite{Klebanov:2000nc} is a singular solution which is dual to the chirally symmetric phase of the Klebanov-Strassler model \cite{Klebanov:2000hb} which has D3-brane charges that dissolve into the flux \cite{Bena:2012ek,PandoZayas:2000ctr}.

The metric is
\begin{gather}
ds_{10}^2=h(r)^{-1/2} \left[ -dt^2+d \vec{x}^2 \right]+h(r)^{1/2} \left[ dr^2+r^2 ds_{T^{1,1} }^2 \right].
\end{gather}

Here $ds_{T^{1,1} }^2$ is a base of a cone with the definition of $ds_{T^{1,1} }^2=\frac{1}{9} (g^5)^2+\frac{1}{6} \sum_{i=1}^4 (g^i)^2$.  It is the metric on the coset space $T^{1,1}=(SU(2) \times SU(2))/U(1)$. Also, $g^i$ are some functions of the angles $\theta_1, \theta_2, \phi_1, \phi_2, \psi$ as
\begin{gather}
g^1=(-\sin \theta_1 d\phi_1-\cos \psi \sin \theta_2 d\phi_2 +\sin \psi d\theta_2)/ \sqrt{2}, \ \ \ \ 
g^2=(d\theta_1-\sin \psi \sin \theta_2 d\phi_2-\cos \psi d\theta_2) / \sqrt{2},\nonumber\\
g^3=(-\sin \theta_1 d\phi_1+\cos \psi \sin \theta_2 d \phi_2-\sin\psi d \theta_2) \sqrt{2}, \ \ \ \ \ 
g^4=(d\theta_1+ \sin \psi \sin \theta_2 d\phi_2+\cos \psi d\theta_2) /\sqrt{2},\nonumber\\
g^5=d\psi+\cos \theta_1 d\phi_1+\cos \theta_2 d\phi_2,
\end{gather}
and also we have the relations $h(r)=\frac{L^4}{r^4} \ln{\frac{r}{r_s}}$ and $L^4=\frac{81}{2} g_s M^2 \epsilon^4$.

In this frame, the asymptotic flat region has been eliminated. Also, $r=r_s$ is where the confining wall is located, which in other metrics has been denoted by $u_{KK}$.

The width of one strip versus turning point of RT surface, and the entanglement entropy of the connected solution in KT background can be written as \cite{Ghodrati:2015rta,Kol:2014nqa}
\begin{gather}
L(r_0)=9\sqrt{2} M \sqrt{g_s} \epsilon^2 \int_{r_0}^\infty dr \frac{\sqrt{\ln \frac{r}{r_s} } }{r^2 \sqrt {\frac{r^6}{r_0^6} \frac{\ln \frac{r}{r_s} }{\ln \frac{r_0}{r_s} }-1 } }, \ \ \
S_C(r_0)= \frac{12 V_2 \pi^3 M^2 g_s \epsilon^4}{G_N^{(10)} } \int_{r_0}^\infty dr \frac{r \ln \frac{r}{r_s} }{\sqrt{1-\frac{r_0^6}{r^6} \frac{\ln \frac{r_0}{r_s} }{\ln \frac{r}{r_s} } }}.
\end{gather}

 \begin{figure}[ht!]
 \centering
  \includegraphics[width=9 cm] {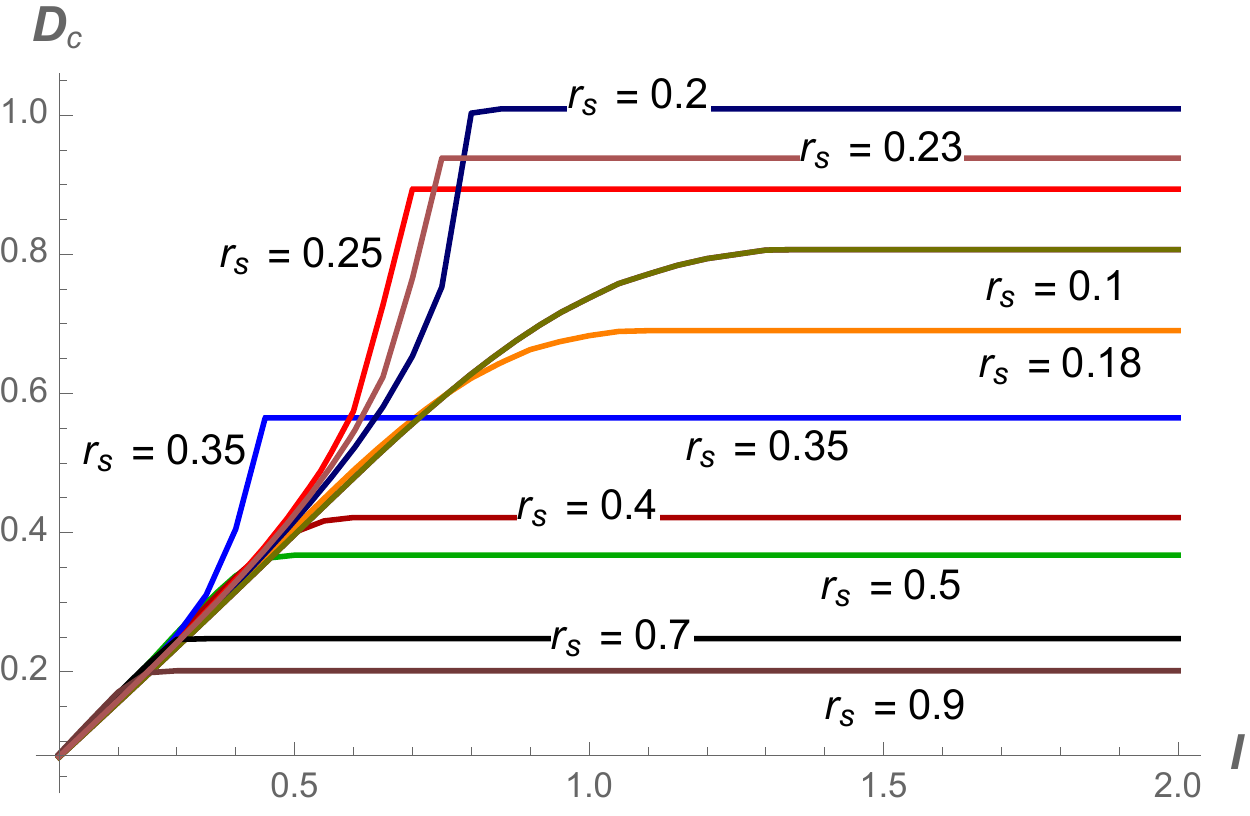} 
  \caption{The relationship between the critical distance $D_c$ between the two strips versus $l$ in the Klebanov-Tseytlin geometry. The regions below each curve for each $r_s$ correspond to non-zero mutual information and $r_s$ corresponds to the position of the IR wall or the cur-off point.}
 \label{fig:regionDKT2}
\end{figure}

Again, using the relations for the entanglement entropy in the confining models, one can calculate the mutual information and then the critical distance between the two strips and then by changing the scales of the system relative to each other and by using $D_c$, one can probe the phase structure in this model where the result is shown in figure \ref{fig:regionDKT2}. From this figure, it could be seen that again four different phases could be detected in these top-down QCD models which is consistent with the result of \cite{Jain:2020rbb}.

\subsection{Maldacena-Nunez}
The Maldacena-Nunez (MN) metric \cite{Maldacena:2000mw,Maldacena:2000yy} is obtained by a large number of D5-branes wrapping on $S^2$ \cite{Maldacena:2000yy}. In the string frame the metric and the fields are written as \cite{Basu:2012ae}
\begin{gather}
ds^2_{10}=e^\phi [ -dt^2+{dx_1}^2+{dx_2}^2+{dx_3}^2+e^{2h(r)}({d\theta_1}^2+\sin^2 \theta_1 {d\phi_1}^2)+dr^2+\frac{1}{4}(w^i-A^i)^2 ], \nonumber\\ 
A^1=-a(r)d\theta_1, \ \ \ \ \ \ \ \ A^2=a(r)\sin \theta_1 d\phi_1, \ \ \ \ \ \ \ \ 
A^3= -\cos \theta_1d\phi_1, \nonumber\\
w^1 =\cos \psi d\theta_2+\sin \psi \sin\theta_2 d \phi_2,\ \ \ \ \ \ \ \ \ 
w^2 =-\sin \psi d \theta_2+\cos \psi \sin \theta_2 d\phi_2,\nonumber\\
w^3=d\psi+\cos \theta_2 d\phi_2,
\end{gather}
and also the other parameters of the metric would be
\begin{gather}
a(r)=\frac{2r}{\sinh 2r}, \ \ \ \ 
e^{2h}= r\coth {2r}-\frac{r^2}{{\sinh 2r}^2}-\frac{1}{4},\ \ \ \ 
e^{-2\phi} =e^{-2\phi_0} \frac{2 e^h}{\sinh 2r}.
\end{gather}

For this geometry we get
\begin{gather}
\alpha(r)= e^\phi,  \ \ \ \ \ \beta(r)=1, \ \ \ \ \ H(r)=64 e^{4h+4\phi}=  64 \left ( r \coth 2r - \frac{r^2}{\sinh 2 r^2} - \frac{1}{4} \right )^2.
\end{gather}

So using the relation for the width of the strip versus turning point, and the entanglement entropy of the connected solution of the strip, for the MN case we get
\begin{gather}
L(r_0)= 2\int_{r_0}^\infty dr \frac{1}{\sqrt{\frac{r \coth 2r- \frac{r^2}{\sinh 2r^2} - \frac{1}{4} }{r_0 \coth 2 r_0 - \frac{r_0^2}{\sinh 2 r_0^2} - \frac{1}{4} }-1} }, \ \ \ 
S_C (r_0) = \frac{4 V_2}{ G_N^{(10)} } \int_{r_0}^\infty \frac{r \coth 2r - \frac{r^2}{\sinh 2 r^2} - \frac{1}{4}  }{\sqrt{1- \frac{r_0 \coth 2 r_0- \frac{r_0^2}{\sinh 2 r_0^2} - \frac{1}{4} }{r \coth 2r - \frac{r^2}{\sinh 2r^2} - \frac{1}{4} } } }.
\end{gather}

\begin{figure}[ht!]
 \centering
  \includegraphics[width=7 cm] {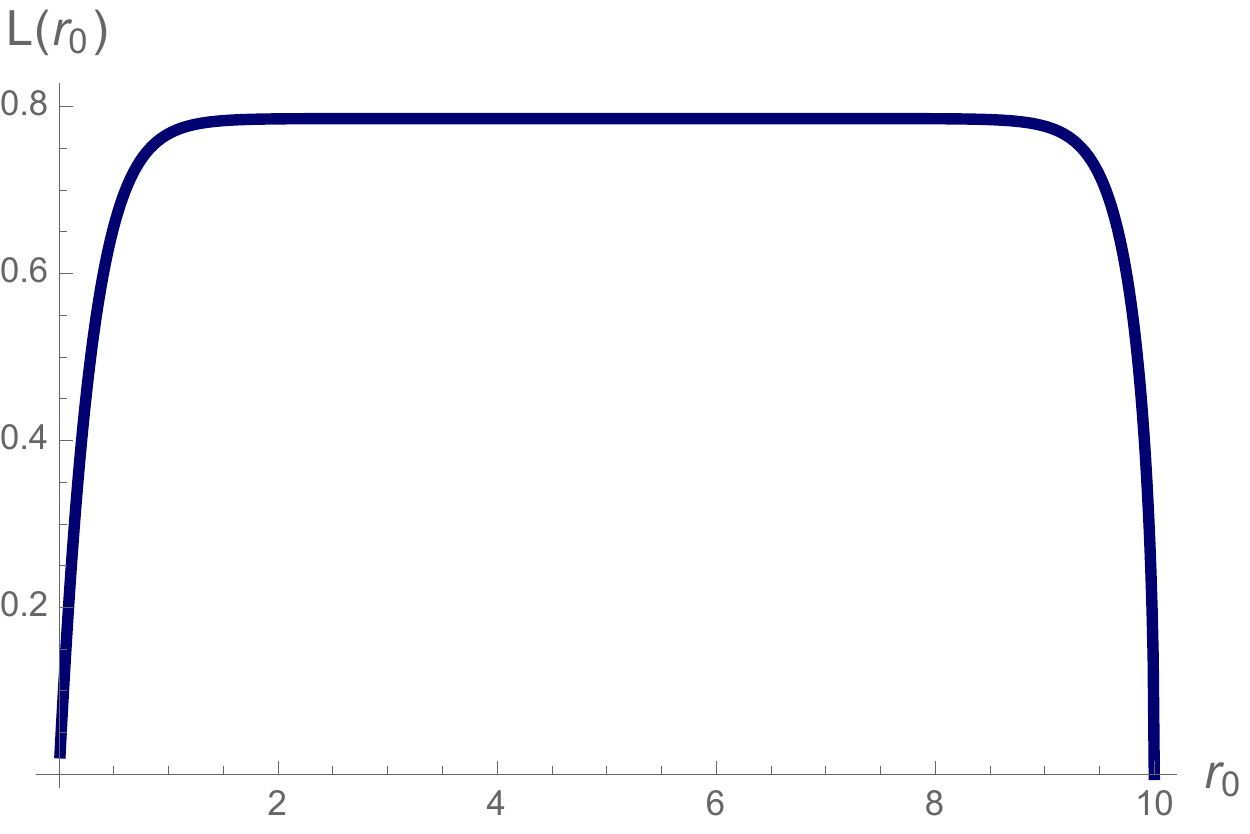} 
    \includegraphics[width=7 cm] {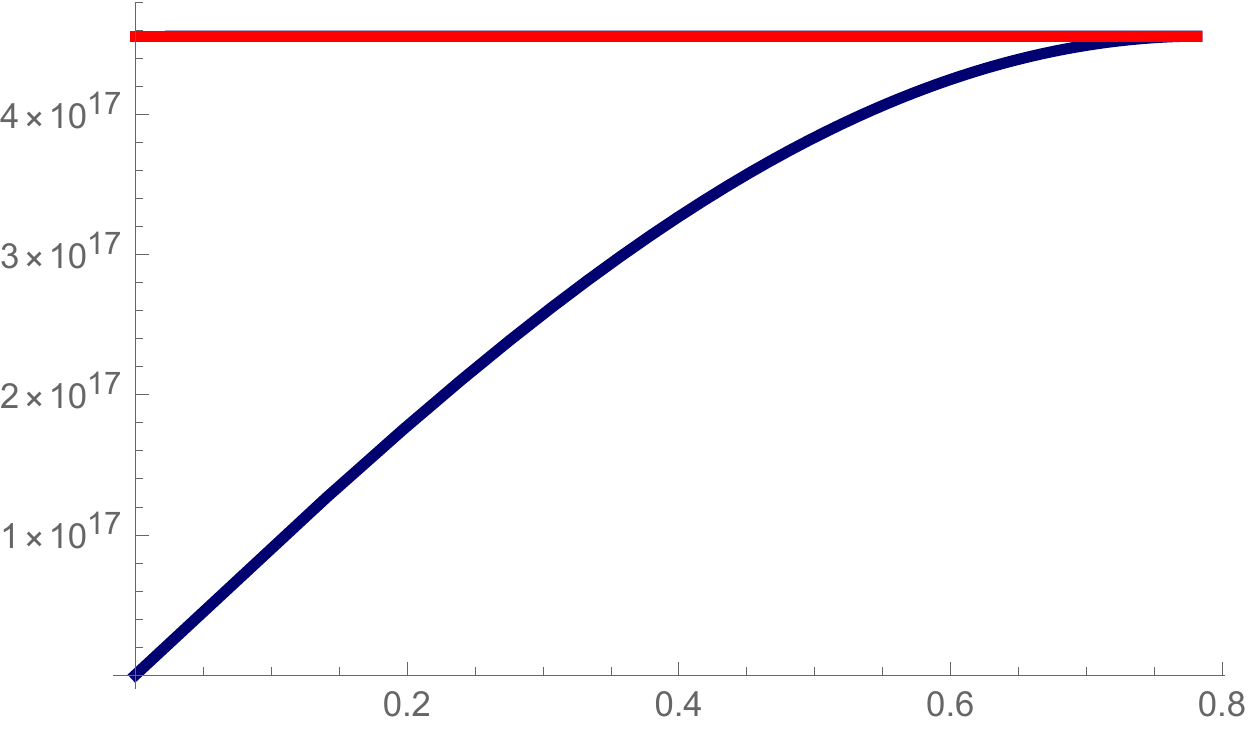} 
  \caption{The relationship between width of a strip versus turning point $r_0$ (in the left), and entanglement entropy versus $L$ (in the right) in the Maldacena-Nunez background.}
 \label{fig:SLinMN}
\end{figure}

The behavior of $L(r_0)$ and $S$ are shown in figure \ref{fig:SLinMN}, where the confining behavior can be detected.

 \begin{figure}[ht!]
 \centering
  \includegraphics[width=7.5 cm] {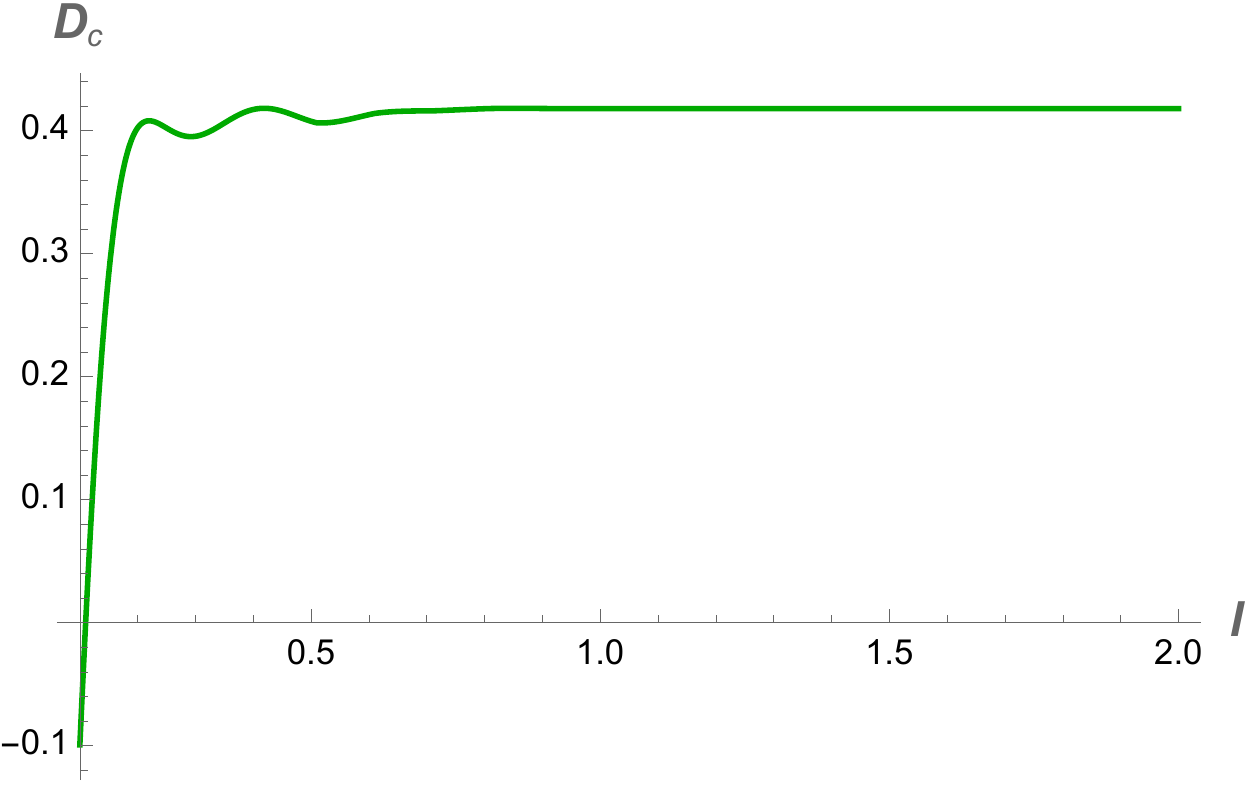} 
  \caption{The behavior of critical distance in the Maldacena-Nunez background.}
 \label{fig:regionDMN1}
\end{figure}

The behavior of the turning point  versus $L$ (one strip) and entanglement entropy are shown in figure \ref{fig:SLinMN} and the behavior of critical distance is shown in figure \ref{fig:regionDMN1}, where again the effects of the non-conformal  background can be seen in the behavior of $D_c$ compared to the previous studies for the conformal cases in \cite{Ghodrati:2019hnn}. In \cite{Ghodrati:2018hss}, we established the connections between complexity and potential in the QCD models. It would be interesting to study $D_c$ in those models as well.

\begin{figure}[ht!]
 \centering
  \includegraphics[width=8.5cm] {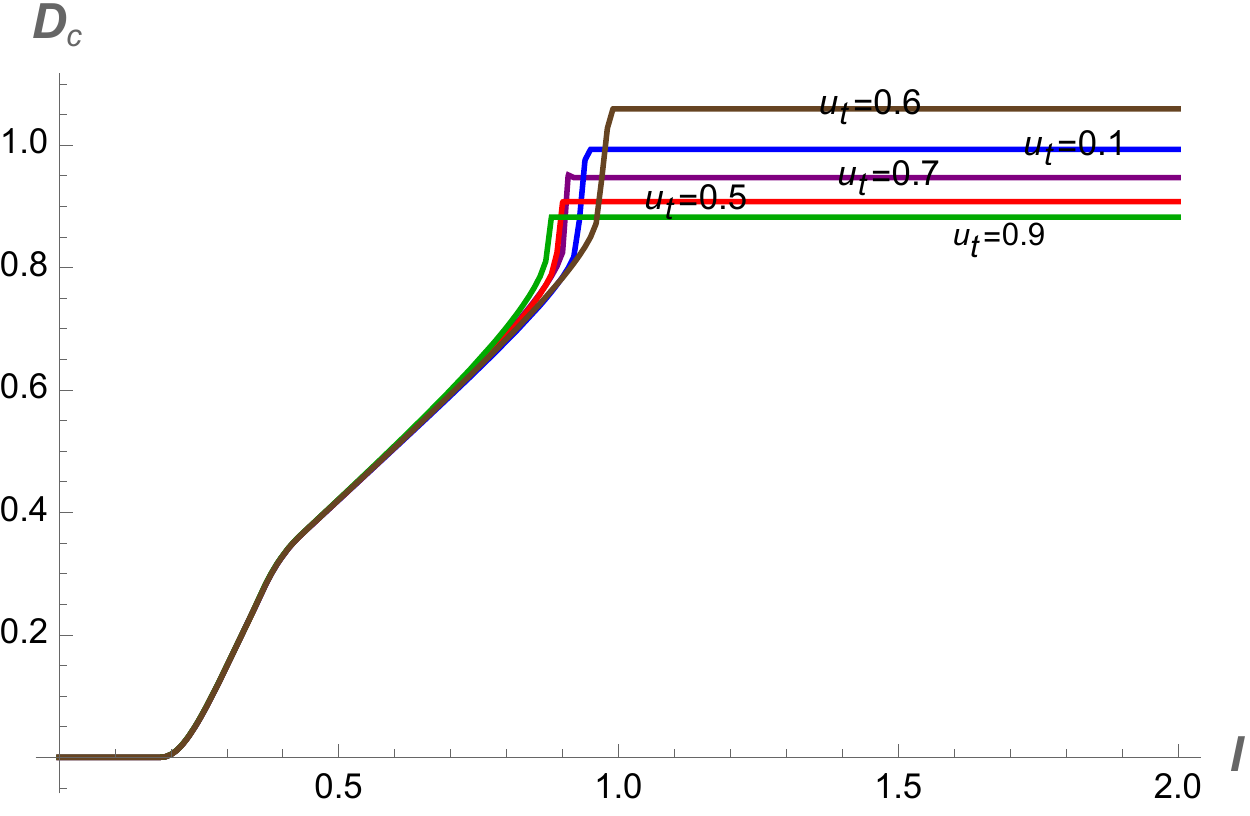} 
  \caption{The phase diagram for $D_c$, coming from $S_C-S_D$. }
 \label{fig:cregularizedphases}
\end{figure}

Note also that for constructing these results we just used the connected part of the entanglement entropy, i.e, $S_C$. We could also use the regularized entropy, $S_C-S_D$, where $S_D$ is the solution for the ``disconnected part'' which is just a constant, and then in the numerical codes we could take the minimum of $S_C$ and $S_D$. For example for the Witten-QCD case, we could get a different phase diagram as in figure \ref{fig:cregularizedphases}. But the general results would be similar and no more phases can be detected using this quantity.

\section{Crofton form in confining backgrounds}\label{sec:Crofton}

In \cite{Czech:2015qta}, using ideas from integral geometry, the authors found the connections between the length of a curve and the number of geodesics, or ``random'' lines it would be intersected and therefore its connection to the Crofton formula and tensor network. The behavior of Crofton form can give further intuitions about the structures of these various confining backgrounds from the perspective of the bulk reconstruction and can shed more lights on their various specific properties. So for checking the interconnections between geometry and the quantum information measures, it would be interesting to calculate the Crofton form for these confining models as well.

\subsection{Sakai-Sugimoto and deformed Sakai-Sugimoto}
The Crofton form of Sakai-Sugimoto geometry would be
\begin{gather}
\omega_{\text{Sakai-Sugimoto} }=\frac{V_3 V_4 R^3_{D_4}  } {2g_s^2 G_N^{(10) } }   \frac{u^4 \Big(10 u_{\text{KK}}^3 u_t^5-5 u^5 u_{\text{KK}}^3-7 u^3 u_t^5+2 u^8\Big)}{2 \Big((u^3-u_{\text{KK}}^3) (u^5-u_t^5)  \Big)^{\frac{3}{2} }}  du \wedge d\theta.
\end{gather}

\begin{figure}[ht!]
 \centering
  \includegraphics[width=8 cm] {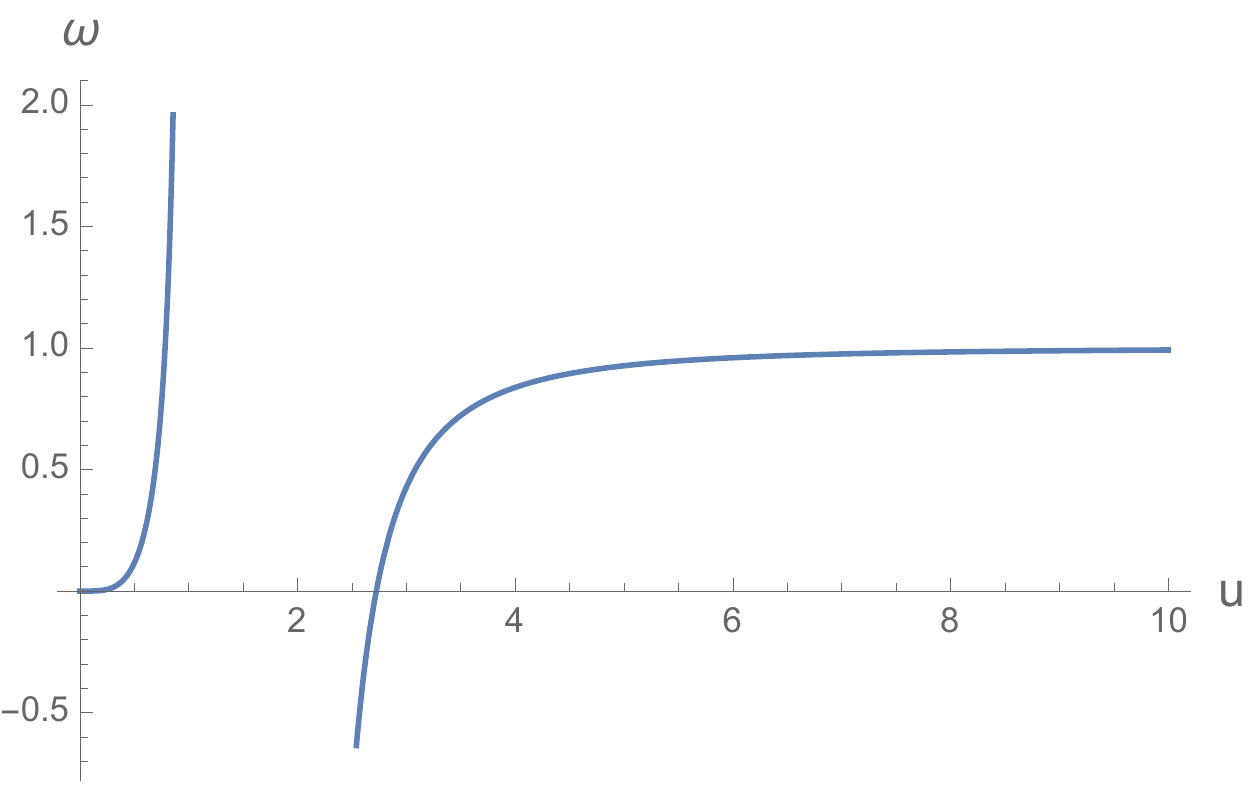} 
  \caption{The plot of Crofton form $\omega$ vs. $r$ in the Sakai-Sugimoto background.}
 \label{fig:croftonSS}
\end{figure}

The plot of Crofton form $\omega$ versus $u$ for $u_{KK}=1$ and $u_t=2$ has been shown in figure \ref{fig:croftonSS}. One can detect a well around $u_t$ and it becomes constant in the large values of $u$ which is a universal behavior in all the confining models that we checked.

Then, using the relation for the deformed case we get
\begin{equation} \label{deformedcroft}
\begin{split}
\omega_{\text{deformed Sakai-Sugimoto} } &=\frac{V_3 V_4 R^3_{D_4}  } {2g_s^2 G_N^{(10) } } \frac{u^4}{2 \left({\frac{u^8-u^5 u_{\text{KK}}^3}{u_0^8-u_0^5 u_{\text{KK}}^3}-1}\right)^{\frac{1}{2}} \left(u^3-u_{\text{KK}}^3\right){}^{\frac{3}{2} } \left(u^5-u_t\right){}^{\frac{3}{2} }}\times  \nonumber\\ &
\Bigg (  \frac{R_{D_4}^{\frac{3}{2}} (3 u_0^8 u^3-(2 u^5+3 u_0^5) u_{\text{KK}}^6+(13 u^8-3 u_0^5 u^3+3 u_0^8) u_{\text{KK}}^3-11 u^{11}) (u^5-u_t^5)}{u_0^5 u_{\text{KK}}^3-u^5 u_{\text{KK}}^3+u^8-u_0^8}  \nonumber\\ &
+R_{D_4}^{\frac{3}{2}} \left(7 u_{\text{KK}}^3 u_t^5-2 u^5 u_{\text{KK}}^3-10 u^3 u_t^5+5 u^8\right)  \nonumber\\ &
+ \Big(\frac{u^8-u^5 u_{\text{KK}}^3}{u_0^8-u_0^5 u_{\text{KK}}^3}-1\Big)^\frac{1}{2} \left(10 u_{\text{KK}}^3 u_t^5-5 u^5 u_{\text{KK}}^3-7 u^3 u_t^5+2 u^8\right)  \Bigg ) du \wedge d\theta,
\end{split}
\end{equation}

\begin{figure}[ht!]
 \centering
  \includegraphics[width=8 cm] {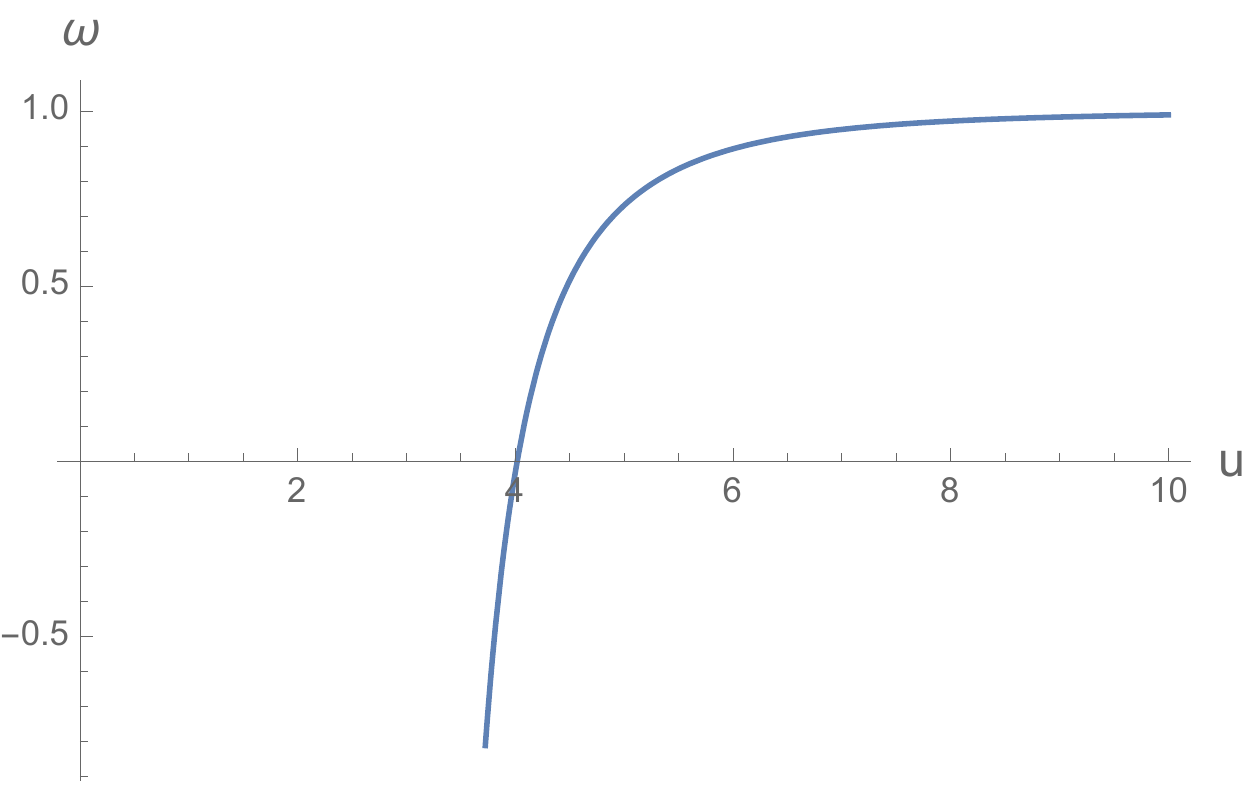} 
  \caption{The plot of Crofton form $\omega$ versus $r$ in the \textit{deformed Sakai-Sugimoto background}. }
 \label{fig:croftondSS}
\end{figure}

The general behavior of  $\omega_{\text{deformed Sakai-Sugimoto} }$ versus $u$ is shown in figure \ref{fig:croftondSS}.

One could see that for the intermediate and bigger values of $u$ the behavior is similar to the Sakai-Sugimoto case, as one would expect.

\subsection{Klebanov-Tseytlin}

The Crofton form for the Klebanov-Tseytlin background would be
\begin{gather}
\omega_{KT}=c_{KT}\frac{2 r^9 \log ^{\frac{3}{2}} (\frac{r}{r_s} )  \big(\log  (\frac{r}{r_s} )+1 \big)-r_0^6 r^3 \log (\frac{r_0}{r_s} ) \log^{\frac{1}{2} }  (\frac{r}{r_s} ) \big(8 \log (\frac{r}{r_s} )+3\big )}{2  \left(r^6 \log  \big(\frac{r}{r_s} \big)-r_0^6 \log  \big (\frac{r_0}{r_s} \big) \right)^{\frac{3}{2} }},
\end{gather}
where $c_{KT}=\frac{12 V_2 \pi^3 M^2 g_s \epsilon^4}{ G_N^{(10)} }$ is just a constant.

\begin{figure}[ht!]
 \centering
  \includegraphics[width=8 cm] {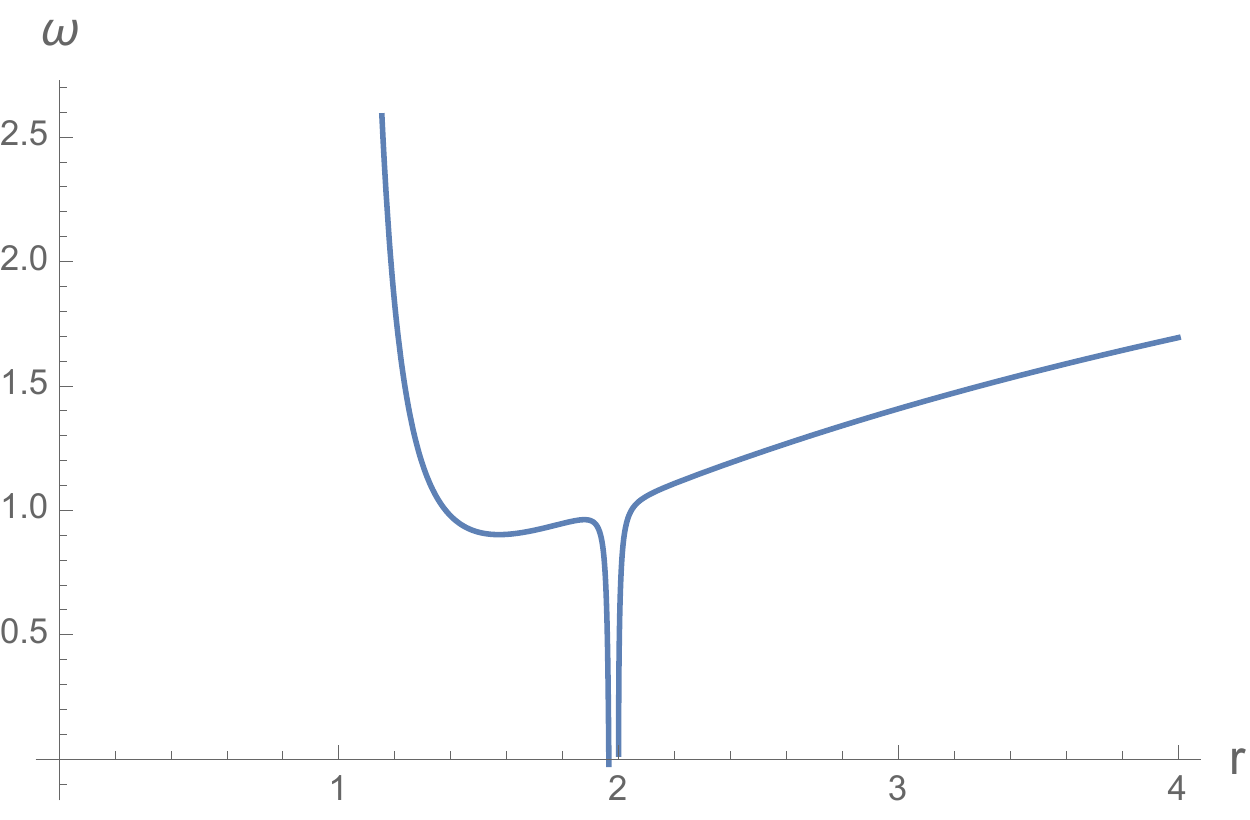} 
  \caption{The plot of Crofton form $\omega$ vs. $r$ in the KT background. }
 \label{fig:croftonKT}
\end{figure}

The plot of $\omega_{KT}$ versus $r$,  for $r_s=2$ and turning point of $r_0=1$ is shown in figure \ref{fig:croftonKT}.

Again, one can see that there is a well around the cut-off point, which signals the fact of the difficulty of bulk reconstruction there and the significant effect it would have on the correlation structures and on phase diagrams.

\subsection{Maldacena-Nunez}

The Crofton form in the Maldacena-Nunez (MN) background would be
\begin{gather}
\omega_{MN}= c_{MN} \frac{2 \sinh (4 r) \left(\sinh ^4(2 r)-2 \sinh ^4\left(2 r_0\right)\right)}{\left(\sinh ^4(2 r)-\sinh ^4\left(2 r_0\right)\right) \sqrt{1-\sinh ^4\left(2 r_0\right) \text{csch}^4(2 r)}},
\end{gather}
where $c_{MN}=\frac{V_2 \pi^3 e^{4\phi_0} }{G_N^{(10)} } $.

\begin{figure}[ht!]
 \centering
  \includegraphics[width=8 cm] {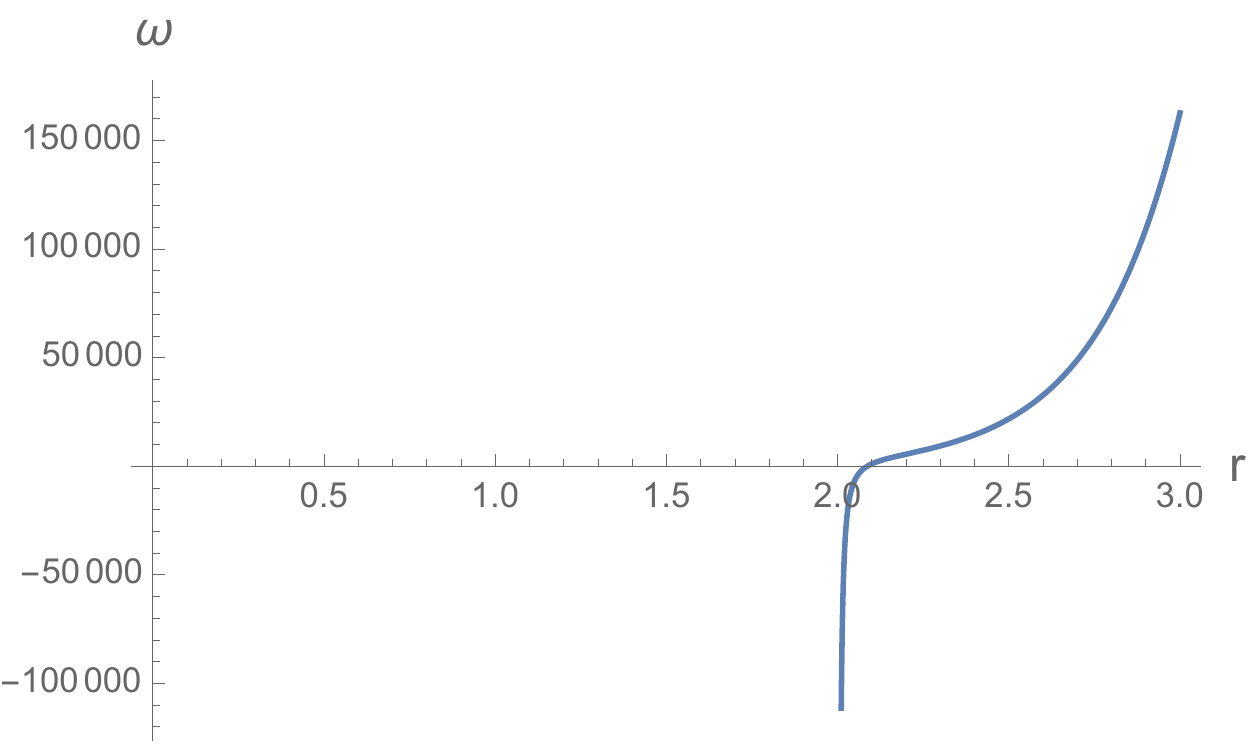} 
  \caption{The plot of Crofton form $\omega$ vs. $r$ in the MN background. }
 \label{fig:croftonMN}
\end{figure}

The plot of Crofton form for $r_0=2$ in the MN case has been shown in figure \ref{fig:croftonMN}. Again, a deep well around the cut-off point could be detected, signaling its effects on the correlations there.

\section{Conclusion}\label{sec:conclude}

In this work, we numerically show that in addition to other quantum information measures, the critical distance between two symmetric strips, $D_c$, where the mutual information drops to zero, as a measure of mixed correlations in the system, could also be a new tool for probing the phase structure of confining models. The phase diagram can be constructed by varying the main three scales in the system, which are the distance between the subsystems $D$, the width of the strips $L$, and the position of the IR wall $r_{\Lambda}$ with respect to each other in the confining models. In the backgrounds where there are both confinement/deconfinement phase transition and the chirality breaking-restoration scale, in addition to entanglement wedge becoming connected or disconnected, the jumps in behavior of the $D_c$ could also be associated to these scales in the theory such as in the case of Sakai-Sugimoto model.

For these confining models, we also present the behavior of Crofton form which is a tool for bulk reconstruction in holography, and so we show that there is a universal behavior around the IR point, as the Crofton form diverges, and also in the large holographic radiuses it becomes constant, signaling the effects of the confining wall on the correlation structures of confining models, and the significant effects it then should have on the phase diagrams, as we showed in the first part of the work.

\section*{Acknowledgments}
 I would like to thank Tuna Demircik for useful discussions. This work has been supported by an appointment to the JRG Program at the APCTP through the Science and Technology Promotion Fund and Lottery Fund of the Korean Government. It has also been supported by the Korean Local Governments – Gyeongsangbuk-do Province and Pohang City – and by the National Research Foundation of Korea (NRF) funded by the Korean government (MSIT) (grant number 2021R1A2C1010834).

\end{document}